\documentclass{revtex4}
\usepackage{booktabs}
\usepackage{titlesec}
\usepackage{graphicx}

\begin{document}

\title{Improving Students¡¯ Understanding of Quantum Measurement\\  Part 2: Development of Research-based Learning Tools}

\author {Guangtian Zhu}
\affiliation {School of Education Science, East China Normal University, Shanghai, China, 200062}

\author {Chandralekha Singh}
\affiliation {Department of Physics \& Astronomy, University of Pittsburgh, Pittsburgh, PA, 15260}

\begin{abstract}
We describe the development and implementation of research-based learning tools such as the Quantum Interactive Learning Tutorials (QuILTs) and peer instruction tools to reduce students' common difficulties with issues related to measurement in quantum mechanics. A preliminary evaluation shows that these learning tools are effective in improving students¡¯ understanding of concepts related to quantum measurement.
\end{abstract}

\maketitle

\section{INTRODUCTION}

\ \ \ \ Issues related to measurement in quantum mechanics are very different from measurements in classical mechanics and students usually struggle in learning about quantum measurement within the standard interpretation. In the first of the two papers (Part 1, Ref.[1]), we describe the investigation of students' common difficulties with quantum measurement within the traditional interpretation which is universally taught to students. Based upon the findings of the investigation, we have developed research-based learning tools to help students build a better knowledge structure about quantum measurement. These research-based learning tools include Quantum Interactive Learning Tutorial (QuILT) and peer-instruction tools such as concept tests similar to those popularized by Mazur for introductory physics courses [2,3]. The QuILT for quantum measurement uses a guided inquiry-based approach [4] to learning and helps students in discerning the coherence in the framework of quantum mechanics related to quantum measurement. It can either be used as an in-class tutorial on which two or three students can work together with full class discussion and instructor feedback as appropriate, or they can be given as homework supplements [5]. The concept tests can be integrated with lectures and encourage students to take advantage of their peers' expertise and learn from each other.

In this paper (Part 2), we will describe the development of the research-based QuILT and concept tests to help students develop a good understanding of quantum measurement within the standard interpretation. We will also discuss preliminary evaluation results of using these research-based learning tools in class. The QuILT and concept tests related to quantum measurement were administered to students in the first semester of a full-year junior-senior level quantum mechanics course. They strive to build on students' prior knowledge, actively engage them in the learning process and help them build links between the abstract formalism and conceptual aspects of quantum physics without compromising the technical content. To assess the effectiveness of the QuILT and concept tests, we administered the same assessment related to quantum measurement to the experimental group and a comparison group in different but equivalent classes at two similar universities. The comparison group only had traditional lectures and weekly homework in a similar two-semester quantum mechanics class in which the same textbook was used. Our prior investigation shows that the students' performance on surveys given in the upper-level quantum mechanics courses at the two universities (experimental group and comparison group) were comparable when traditional instruction was used at both institutions. We find that students who use research-based learning tools perform significantly better than those who do not use them. Below, we elaborate on the research-based learning tools that the students used to learn about quantum measurement.

\section{PEER INSTRUCTION AND CONCEPT TESTS}

\ \ \ \ \ In the peer instruction approach, students reflect with peers upon problems. Integration of peer interaction with lectures has been popularized in the physics community by Mazur [2]. In Mazur's peer interaction approach, the instructor poses conceptual problems or concept tests in the form of multiple-choice questions to students periodically during the lecture. The focal point of the PI method is the discussion among students, which is based on conceptual questions; the lecture component is limited and intended to supplement the self-directed learning. The conceptual multiple choice questions give students an opportunity to think about the physics concepts and principles covered in the lecture and discuss their answers and reasoning with peers. The instructor polls the class after peer interaction to obtain the fraction of students with the correct answer. Students learn about the level of understanding that is desired by the instructor by discussing with each other the concrete questions that are posed as concept tests. The feedback obtained by the instructor is also invaluable because the instructor learns about the fraction of the class that has understood the concepts at the desired level. This peer instruction strategy keeps students alert during lectures and helps them monitor their learning, because not only do they have to answer the questions, they must explain their answers to their peers. The method keeps students actively engaged in the learning process and lets them take advantage of each others' strengths. It also helps high-performing students, because explaining and discussing concepts with peers helps students organize and solidify concepts in their minds.

Our prior research has shown that, even with minimal guidance from the instructors, students can benefit from peer interaction [6]. In our study, those who worked with peers not only outperformed an equivalent group of students who worked alone on the same task, but collaboration with a peer led to ¡°co-construction¡± of knowledge. Co-construction of knowledge occurs when neither student who engaged in the peer collaboration was able to answer the questions before the collaboration, but both were able to answer them after working with a peer on a post-test given individually to each person. For example, in our prior research [6], introductory physics students first answered the questions in the Conceptual Survey of Electricity and Magnetism (CSEM) [7] individually after traditional instruction. Then, they paired up and discussed the questions with their partners and answered the questions again in pairs. The fraction of responses on each question that went from both incorrect individually to the correct response from the pair is 29\% and shows evidence for co-construction. Individual discussions suggest that students benefited from discussing their doubts with their peers [6].

The classroom at the University of Pittsburgh (Pitt) in which quantum mechanics 1 was taught was equipped with a clicker system so that students could submit their answers electronically. Students were actively engaged in the peer discussion. The distribution of answers was displayed to the whole class after all the students had made their choices following discussion with peers. The professor provided further explanations based upon students' responses.

In the concept tests related to quantum measurement or its pre-requisite, we designed a sequence of multiple-choice questions to address similar concepts in different contexts [8]. For example, some concept tests dealt explicitly with how the identity operator can be written as a complete set of eigenstates of an operator corresponding to physical observable with discrete or continuous eigenvalues (e.g., $ \sum\limits_{n}|n \rangle \langle n|  = \hat{I} $ \ or $ \int\limits_{all}^{}| x \rangle \langle x|dx = \hat{I} $ ). Students learned how to write any state of the system in terms of a complete set of eigenstates by using this identity operator. They also learned about calculating the probability amplitude for measuring a particular value for an observable by projecting the state along the corresponding eigenstate of the operator as in the following concept test question:
\begin{itemize}
   \item   \emph{Suppose $  |\psi \rangle$ is a vector in the Hilbert space which represents the state of the system at time t=0.
$  |n \rangle$ are the eigenstates of the Hamiltonian operator \^{H}  with eigenvalues $E_{n}$ .Choose all of the following statements that are correct.}
\end{itemize}
\emph{(1) $  |\psi \rangle $ = $ \sum\limits_{n}|n \rangle \langle n| \psi \rangle $  \\
(2) $  e^{-i \hat{H} t \// \hbar} |\psi \rangle = \sum\limits_{n} e^{-i E_{n} t \// \hbar} |n \rangle \langle n| \psi \rangle$ \\
(3) If we measure the energy of the particle in the state $|\psi \rangle$ , the probability of obtaining $E_{n}$ is $ |\langle     n | \psi \rangle |^2$.}

$A$.1 only \ \ \ \ \ \ \ \ \ \ \ \ \ \ \ \ \ \ $B$.1 and 2 only \ \ \ \ \ \ \ \ \  $C$.1 and 3 only\

$D$.2 and 3 only \ \ \ \ \ \ \ \ \  $E$.all of the above
\vspace{12 pt}

All of the options in this question are correct. Option (2) reviews the necessity of writing the states as a linear superposition of the energy eigenstates in order to determine the time evolution of the state.  Option (3) helps students consider the relationship between the probability of obtaining an energy eigenvalue and projecting the state vector along the corresponding energy eigenstate.

The next concept test question helps students to review similar issues by writing the state vector as a linear superposition of a complete set of eigenstates of position or momentum operators.  In this case, the eigenvalue spectrum is continuous. After answering the previous concept test question, students know that for a complete set of eigenstates with discrete eigenvalues,  $\sum\limits_{n}|n\rangle \langle n|=\hat{I}$, so we ask them to generalize it to the continuous eigenvalue spectrum cases which is a natural extension via concept test questions like the following:
\begin{itemize}
  \item \emph{$|\psi \rangle$ is a vector in the Hilbert space which denotes the state of a quantum particle at time t=0. $| x \rangle$ and $| p \rangle $ are the eigenstates of position and momentum operators. Choose all of the following statements that are correct.}
\end{itemize}
\emph{(1) $|\psi \rangle =\int |p \rangle \langle p|\psi \rangle dp$\\
(2) $|\psi \rangle =\int \psi(x) |x\rangle dx $\\
(3) If we measure the position of the particle in the state $|\psi\rangle$, the probability of finding the particle between x and x+dx is $ |\langle x |\psi \rangle |^2 dx $}

$A$. 1 only \ \ \  $B$. 1 and 2 only \ \ \   $C$.1 and 3 only \ \ \

$D$. 2 and 3 only \ \ \   $E$. all of the above
\vspace{12 pt}

The correct answer to the question above is \emph{E}. As noted earlier, in this question, we changed the context from the energy eigenstates to the position and momentum eigenstates. Students must think about how to replace a complete set of eigenstates with discrete eigenvalues as the basis vectors with a complete set of eigenstates with a continuous spectrum of eigenvalues. Options (1) and (2) can also help reinforce the wavefunction in the momentum representation or the position representation. Option (2) reminds students that the wavefunction in the position representation is the projection of the state $ |\psi\rangle$
onto the position eigenstates, i.e., $\psi(x)=\langle x| \psi  \rangle $. Option (3) helps students consider the probability of position measurement in analogy with the probability of energy measurement in the previous question. The comparison between the two questions can help students understand that for issues related to the measurement of an observable, it is useful to expand the state of the system as a complete set of eigenstates of the corresponding operator and then the absolute square of the expansion coefficient is related to the probability of measurement. Another concept test related to measurement given to the students was the following:

\begin{itemize}
\item \emph{An operator \^{Q} corresponding to a physical observable Q has a continuous non-degenerate spectrum of eigenvalues. $|\psi_q\rangle$  are eigenvectors of \^{Q}  with eigenvalues $q$.  At time t=0, the state of the system is $|\Psi \rangle$ .
Choose all of the following statements that are correct.}
\end{itemize}
\emph{(1) A measurement of the observable Q must return one of the eigenvalues of the operator \^{Q}.\\
(2) If we measure Q at time t=0, the probability of obtaining an outcome between $q$ and q+dq is $|\langle\psi_q|\Psi\rangle|^2 dq$.\\
(3) If we measure Q at time t=0, the probability of obtaining an outcome between $q$ and q+dq is $|\int\limits^{+\infty}_{-\infty}
\psi_{q}^{\ast}(x) \Psi(x) dx |^2 dq $ in which $ \psi_q (x) $ and $ \Psi(x) $  are the wavefunctions corresponding to states $ | \psi_q \rangle $ and $|\Psi \rangle $ respectively.
}

$A$. 1 only \ \ \ \ \ \ \ \ \ \ \ \ \ \ \ \ \ \ $B$. 1 and 2 only \ \ \ \ \ \ \ \ \  $C$. 1 and 3 only\

$D$. 2 and 3 only \ \ \ \ \ \ \ \ \  $E$. all of the above
\vspace{12 pt}

As can be seen, this concept test question (with correct answer \emph{E}) generalizes what students had learned in the previous ones. Note that although all three questions listed above have the answer ``all of the above'', this is not the case for all of the peer instruction questions. These three questions were part of a sequence of concept tests given to students on quantum measurement. The concept tests were used throughout the semester by the students in the experimental group. We use these three concept test questions here to illustrate our strategies for designing the concept tests as a sequence. The students' understanding of quantum measurement in the experimental group partly relied on the effectiveness of using concept tests as a peer discussion tool.

\section{QUANTUM INTERACTIVE LEARNING TUTORIAL (QuILT) RELATED  TO  MEASUREMENT}

\ \ \ \ The goal of the measurement QuILT is to build connections between the formalism and conceptual aspects of quantum measurement without compromising the technical aspects [5]. The measurement QuILT can be implemented in class so that two or three students work together on it. Or it can also be given to the students as homework or self-learning materials [5,9].

The measurement QuILT builds on students' prior knowledge and was developed by taking into account the difficulties found in the written surveys and interviews. QuILT development went through a cyclical iterative process which includes the following stages: (1) Development of the preliminary version based upon theoretical analysis of the underlying knowledge structure and research on students\'\ difficulties, (2) Implementation and evaluation of the QuILT by administering it individually to students, measuring its impact on student learning and assessing what difficulties remained, (3) refinement and modification based upon the feedback from the implementation and evaluation.

Individual interviews with students were carried out using a think-aloud protocol [10] to better understand the rationale for their responses before, during and after the development of different versions of the QuILT and the corresponding pre-test and post-test. During the semi-structured interviews, students were asked to verbalize their thought processes while they answered questions about measurement either as separate questions before the preliminary version of the QuILT was developed or as a part of the QuILT. Students were not interrupted unless they remained quiet for a while. In the end, we asked them for clarification of the issues they had not made clear earlier. Some of these interviews involved asking students to predict what should happen in a particular situation, having them observe what happens in a simulation, and asking them to reconcile the differences between their prediction and observation. After each individual interview with a particular version of the measurement QuILT (along with the pre-test and post-test administered), modifications were made based upon the feedback obtained from the test results and students' performance on the QuILT (if students got stuck at a particular point and could not make progress from one question to the next with the hints already provided, suitable modifications were made). When we found that the measurement QuILT was working well in individual administration and the post-test performance was significantly improved compared to the pre-test performance, it was administered in the quantum mechanics class.

The measurement QuILT uses computer-based visualization tools to help students build a physical intuition about concepts related to quantum measurement [11-15]. The Open Source Physics program [16] was adapted as needed throughout the measurement QuILT. This program is flexible and can be easily tailored to the desired situations. In the measurement QuILT, after predicting what they expect in various situations, students are asked to check their predictions using simulations. If the prediction and observations do not match, students reach a state of disequilibrium and themselves realize that there is some inconsistency in their reasoning. At that point the QuILT provides them appropriate guidance and support to help build a good grasp of relevant concepts and reconcile the difference between their predictions and observations.
\\
\textbf{A. Warm-up exercises for the measurement QuILT }
\\

The measurement QuILT begins with warm-up exercises that students work on before learning from the QuILT. In our research we found that some students have difficulties in understanding the basic concepts about eigenstates of an operator corresponding to a physical observable. Therefore, we designed the warm-up to help students review the concept of eigenstate and to help them understand that the eigenstates of all physical observables are not the same. First, we let students differentiate the energy eigenstates and a possible wavefunction which was a linear superposition of the energy eigenstates. Questions were also designed to help students understand that energy eigenstates satisfy the time independent Schroedinger equation (TISE) but their linear superpositions with different energies do not. In addition to the questions in verbal and mathematical representations that asked students to consider the differences between the energy eigenstates and their linear superposition, one question asked them to select the energy eigenstates from three pictorial representations as shown in Fig 1 (in which the first two were sinusoidal) for a 1D infinite square well.

\vspace{12 pt}
\vspace{12 pt}

\begin{figure}[h]
  \centering
  \includegraphics[width=1\textwidth]{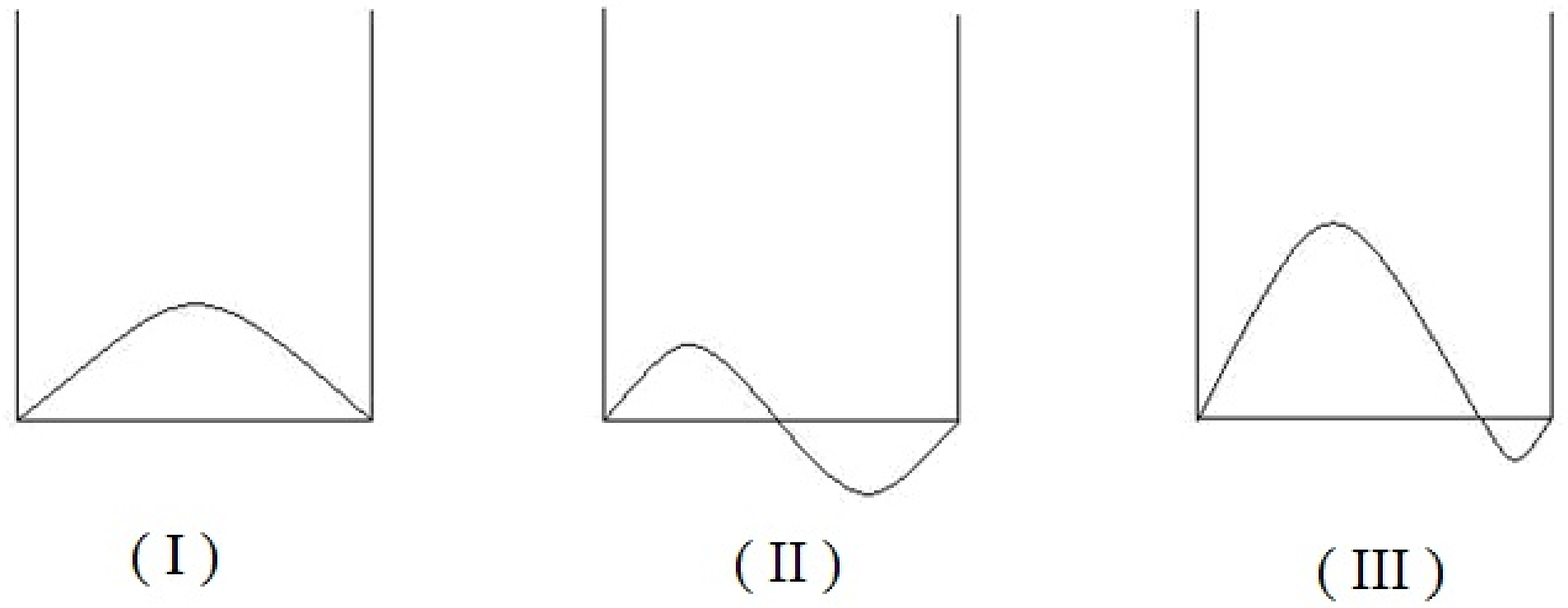}
\end{figure}
\textbf{Fig 1}. Pictorial question in the warm-up testing students' understanding of the energy eigenstates for a 1D infinite square well. (I) and (II) are energy eigenstates but their superposition (III) is not.
\vspace{12 pt}

Pictures (I) and (II) in Fig 1 correspond to the ground and first excited state wavefunctions $\psi_1$ and $\psi_2 $ respectively. Picture (III) is one particular linear superposition of (I) and (II) (e.g., $\psi_1+\psi_2$ ). The warm-up tutorial helps the students learn that the energy eigenstates for this system are even or odd about the center of the well but their superposition need not be. After the 1D infinite square well model, similar considerations were reinforced using the simple harmonic oscillator (SHO) model. From these two models, students learned that the eigenfunctions of different Hamiltonians have different shapes in position space but they satisfy the TISE for the respective systems because they are states with definite energy. Students were required to summarize these characteristics of the energy eigenstates after they studied these two examples in the warm-up.

The position eigenstate was also important in helping students understand the concept of an eigenstate and the fact that not all eigenstates are energy eigenstates. Students were asked to draw a position eigenfunction in the position space with an eigenvalue $x_0$ or a particle interacting with an infinite square well or a finite square well. The warm-up helps students recognize that unlike the energy eigenfunctions, the position eigenfunctions have the same shape for all the 1D systems and their shape has nothing to do with the Hamiltonian of the system. In the warm-up, students also learned about the mathematical representation of a position eigenfunction as a delta function in position space and they were explicitly asked to compare the position eigenfunction and the energy eigenfunction in position space. In one question, they were asked to consider the following statement and explain why they agreed or disagreed:
\begin{itemize}
    \item \emph{``The position eigenstate and energy eigenstate are the same for a given system. After all, they are all eigenstates.'' Explain why you agree or disagree with this statement.}
\end{itemize}

The warm-up helped students learn about the properties of eigenstates of the operators corresponding to different physical observables. Students learned that eigenstates of different operators are different and they satisfy an eigenvalue equation for that operator. They also learned that if the system is in an eigenstate of an operator corresponding to a physical observable, that observable is well-defined in that state and its measurement will yield a definite value with 100\% probability.\\
\\
\textbf{B. Outcome of quantum measurement}
\\

The main measurement QuILT was divided into two sections. One deals with outcomes of measurement and the probability of obtaining those outcomes whereas the other deals with time-evolution after the measurement. The measurement QuILT begins with the basic model of a 1D infinite square well. For different states $|\psi_1 \rangle $, $ \frac{1}{\sqrt2}(|\psi_1 \rangle + | \psi_2\rangle ) $ and $ |\Psi\rangle=\sum A_n|\psi_n\rangle $, students predict what value they would obtain and what state the system would be in after the measurement. After their prediction, they use a computer simulation (adapted from the open source physics simulations) to examine their responses. If a student's prediction is inconsistent with what he/she observes in the simulation, there is a cognitive conflict which provides motivation to resolve the inconsistency [18]. Then the QuILT provides guidance to students to help them reconcile the differences between their predictions and observations so they can build a robust knowledge structure.

\begin{figure}[h]
  \centering
  \includegraphics[width=1\textwidth]{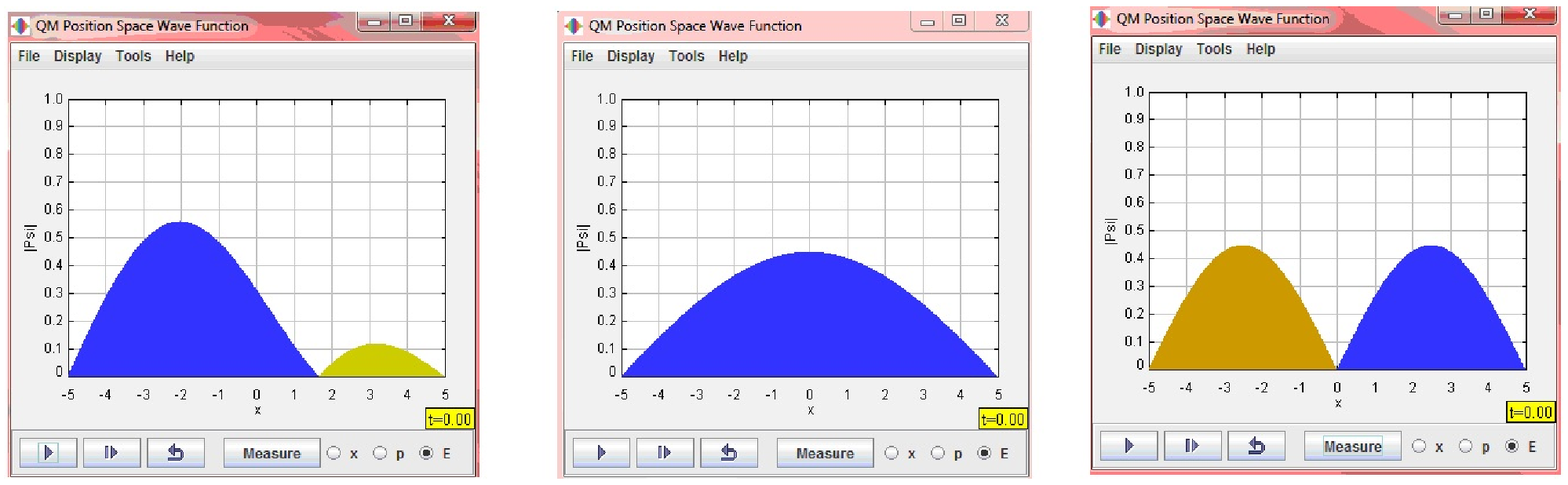}
\end{figure}

\textbf{Fig 2}. Simulation program of the energy measurement on a superposition state (as shown in (a)) with two energy eigenstate components (as shown in (b) and (c)). The vertical axis is the absolute value of the wavefunction (not the probability density for position measurement) and the horizontal axis is the position $x$.\\

In the simulation, one example of an initial state is $(|\psi_1 \rangle +|\psi_2\rangle)\// \sqrt2 $ whose absolute value in position space is shown in Fig 2(a). The vertical axis is the absolute value of the wavefunction and the horizontal axis is the position. Our research of students' difficulties showed that some students mistakenly thought they may obtain the value $ (E_1+E_2)\//2 $ if they measure energy in the superposition of the energy eigenstates $(|\psi_1 \rangle +|\psi_2\rangle)\// \sqrt2 $. In the simulation, students can measure the physical observables of position, momentum and energy to examine the possible outcomes. In \textbf{Fig 2(a)}, students can observe the shape of the absolute value of the superposition state $(|\psi_1 \rangle +|\psi_2\rangle)\// \sqrt2 $ in position space at time t=0. When the students measure the energy of the system, the state of the system may collapse to the energy eigenstates $ |\psi_1 \rangle $ or $ |\psi_2\rangle $ whose absolute values in position space are shown in \textbf{Fig 2(b)} and \textbf{(c)} respectively.

The students are also asked to reset the initial state and repeat the measurement process several times to check whether the measurement yields the same result (the probability is 50\% for obtaining $ |\psi_1 \rangle $ or $ |\psi_2\rangle $). Since the state is a superposition of only two stationary states, it is possible for the students to obtain the same state after the energy measurements. Therefore, the QuILT asked students what could happen if they measured energy in the state $ \sum A_n | \psi_n \rangle $ whose absolute value in position space is shown in Fig 3, and which is a linear superposition of nine stationary states $ |\psi_1\rangle $ or $ |\psi_9\rangle $ with equal probability. After predicting the probability of obtaining different values of energy, students were asked to measure the energy, reset the system to the initial state and measure it again. Since the probability of measuring the same energy is small for this superposition, students appreciated this example while working on the QuILT. To ensure that the students understood the issues related to the energy measurement in multiple contexts, the QuILT also incorporated questions for the SHO Hamiltonian.

\begin{figure}[h]
  \centering
  \includegraphics[width=1\textwidth]{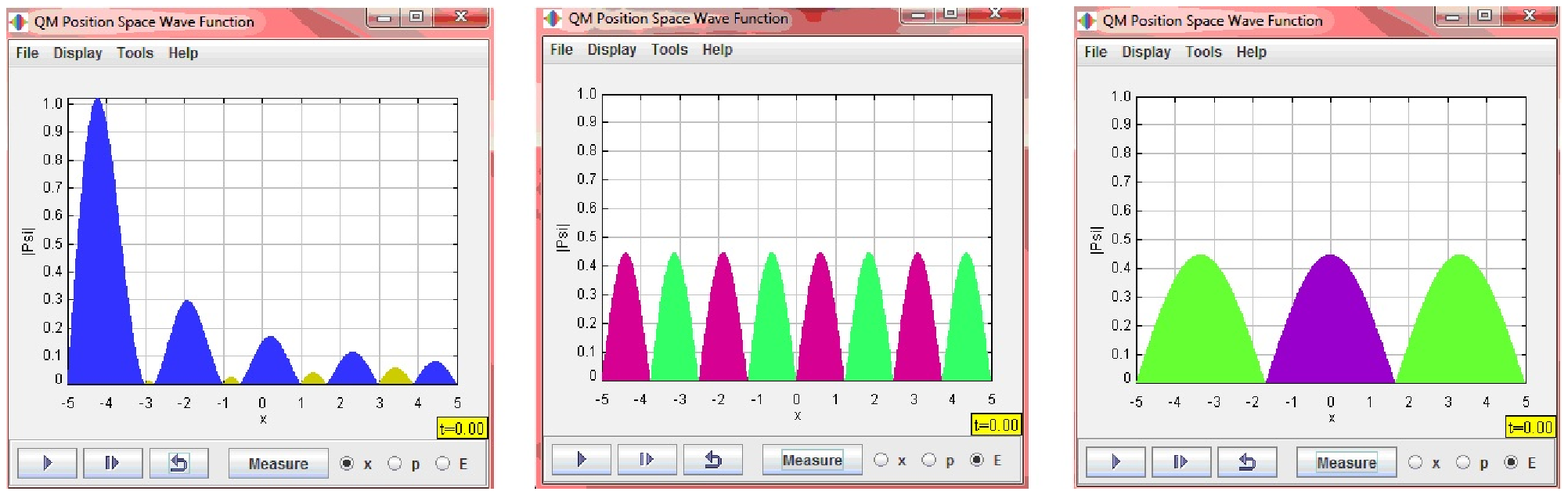}
\end{figure}

\textbf{Fig 3.} Simulation program of the energy measurement on a superposition state (as shown in (a)) with nine energy eigenstate components. (b) and (c) are the examples of absolute values in position space of two basis energy eigenstates of the superposition state.\\

Since the students have difficulties in differentiating between the energy eigenstates and the eigenstates corresponding to other operators corresponding to physical observables, the measurement QuILT also helps students with issues related to the position measurement with initial states similar to those for the energy measurement, e.g., 1D infinite square well and SHO with the initial states $|\psi_1 \rangle $ or $\frac{1}{\sqrt2}(|\psi_1\rangle+|\psi_2\rangle) $. Students first predict theoretically what state they would obtain after a position measurement and then they use the simulation to check their prediction. In an ideal position measurement, the state of the system would collapse to a delta function in position space at a position where the probability of measuring the position is non-zero. As shown in the \textbf{Fig 4}, the initial state $ |\psi_1\rangle $ collapses to a broad peaked Gaussian packet (absolute value shown in position space) because of the computational limitations in constructing a very peaked function. However, the QuILT uses this opportunity to help students recognize that a delta function is a theoretical construction and the position measurement in real world situations, e.g., single particles in double slit experiment landing on the screen, would have an uncertainty in position measured.

  \includegraphics[width=1\textwidth]{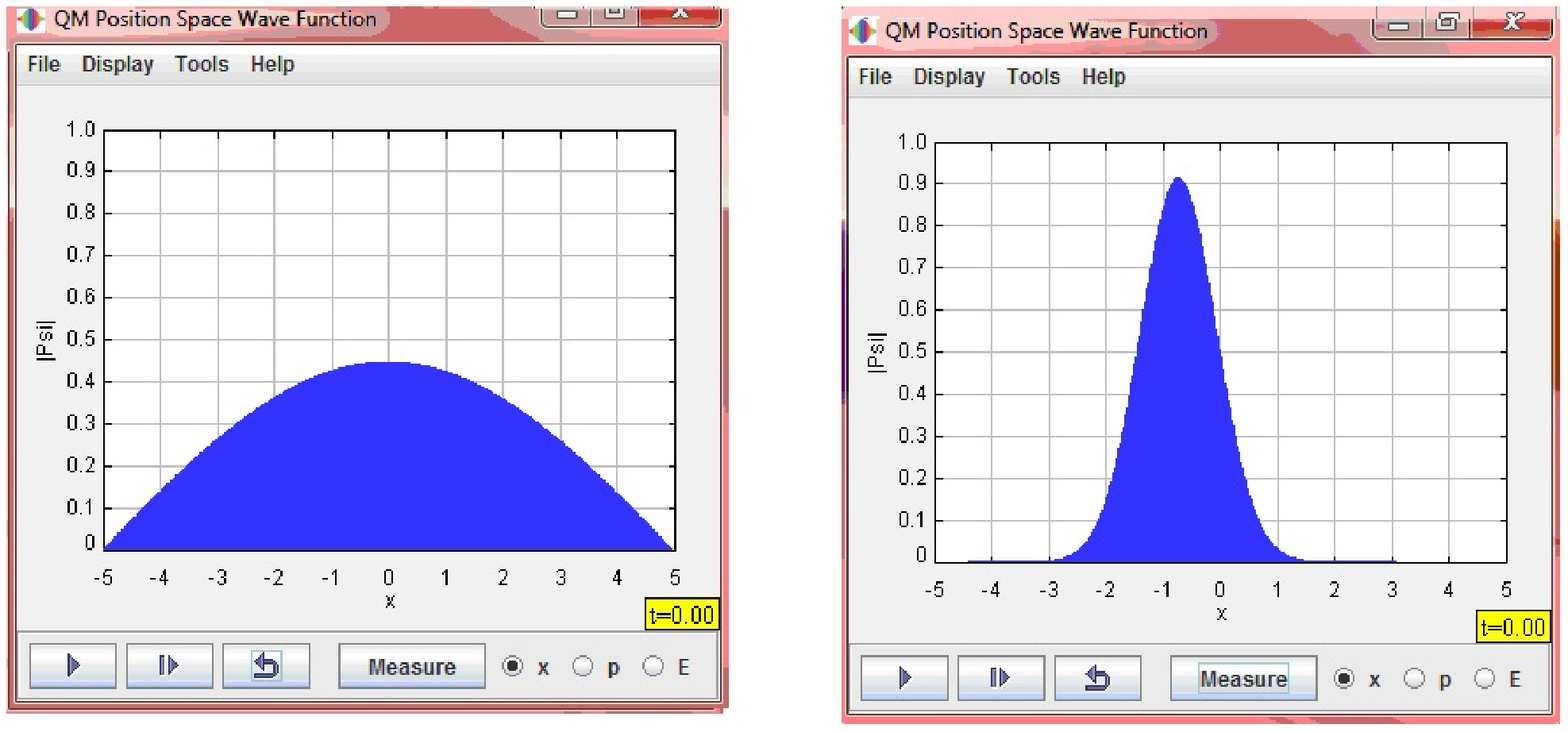}

\textbf{Fig 4}. Simulation program showing one possible outcome of the position measurement on an energy eigenstate (in position space).\\

After predicting what should happen if they perform position measurements on a large number of identically prepared systems, students are asked to reset the initial state of the system and repeat the position measurement. They observe that the center of the collapsed wavefunction is generally different but its shape is always the same. This notion is verified by the students in multiple contexts, e.g., for different quantum systems and different initial states. Students are explicitly asked to compare and contrast what they learned from the measurements of position and energy to help them understand better the outcomes of measurement for different physical observables.\\
\\
\textbf{C. Calculating the Probability of Measuring Different Values}
\\

In addition to helping students learn about the possible outcomes of a measurement, the QuILT also teaches students how to calculate the probability of obtaining each outcome, which is a common difficulty for both undergraduate and graduate students as shown in our research. In surveys and individual interviews, we found that most students could find the probability of measuring different energies by observing the coefficients in an explicit superposition of stationary states, e.g.,  $ \frac{1}{\sqrt{2}}(|\psi_1\rangle+|\psi_2\rangle) $. In the QuILT, students first learn to interpret these coefficients by using the projection of the initial state along an eigenstate of the operator corresponding to the observable measured. In a guided approach, students learn to calculate the coefficients of different energy eigenstates for cases where the wavefunction may not be explicitly written as a linear superposition of stationary states. The QuILT also helps students make connection between the Dirac notation form and integral form of the inner product $\langle\psi_n|\Psi\rangle=\int\limits \psi ^\ast_n (x)\Psi(x)dx$ (a common difficulty with the position representation is that students do not realize that there is an integral involved in writing $\langle\psi_n|\Psi\rangle$ in position space). Students are asked to infer the dimension (unit) of the inner product $ \langle\psi_n|\Psi\rangle $ and the physical meaning of $|\langle\psi_n|\Psi\rangle|^2 $. These abstract inner products are calculated in concrete contexts, e.g., $|\Psi\rangle=\frac{1}{\sqrt{2}} (|\psi_1\rangle+| \psi_2\rangle)$. Students learn that for this concrete case, for $n\geq3$, the probability of obtaining energy $E_n$ is zero because the projection of the state $ |\Psi\rangle $ along the eigenstate $ |\psi_n\rangle $ is zero, i.e., $\langle \psi_n|\Psi\rangle=0$.
 After making sense of the probability for measuring energy for state $|\psi\rangle=\frac{1}{\sqrt{2}}(|\psi_1 \rangle + | \psi_2\rangle)$, students calculate the probabilities of measuring different energies for a general state $|\Psi\rangle=\sum \limits_{n} A_n |\psi_n\rangle $ which is explicitly written as a linear superposition of stationary states. They can find that
$A_n=\langle\psi_n|\Psi\rangle $ is the probability amplitude and $ |\langle\psi_n|\Psi\rangle |^2$  is the probability of measuring energy  $E_n$.

The QuILT then helps the students to understand that any possible state $|\Psi\rangle$ which is not explicitly written as a linear superposition of a complete set of eigenstates of an operator corresponding to a physical observable, e.g., energy, could be written that way. For example, students are asked the following question.
\begin{itemize}
   \item The orthonormal energy eigenfunctions $\psi_n(x)$ for a 1D infinite square well satisfy $\int\limits^{+\infty}_{-\infty} \psi ^\ast_n (x)\psi_m(x)dx=\delta_{mn}$, where $\delta_{mn}=1$ when $m=n$, and $\delta_{mn}=0 $ otherwise. Any state $|\Psi\rangle$ can be expressed as $|\Psi\rangle=\sum \limits_{n} A_n |\psi_n\rangle $ because $|\psi_n\rangle $ form a complete set of vectors for the Hilbert space in which the state of the system lies. Find $A_n$ in terms of $|\Psi\rangle$ and $|\psi_n\rangle $ first in the Dirac notation form and then in the integral form in the position representation.
\end{itemize}
If the students did not have the mathematical skills to answer the question above, hints were provided, e.g., about how to use the Fourier trick and multiply both sides of the expression $\Psi(x)=\sum A_n \psi_n(x)$ by $\psi_m^\ast(x)$and integrate over all space. Then students calculated the probability of obtaining $E_n$ for a concrete example of a triangle-shaped wavefunction for a 1D infinite square well for which the wavefunction was not explicitly written in terms of a linear superposition of energy eigenfunctions. Students further contemplated over these issues in multiple contexts such as the SHO model.

The QuILT helps students learn that the probability of measuring other physical observables can be obtained by projecting the state of the system along an eigenstate of an operator corresponding to a physical observable. They use this projection method to analyze the probability density for position measurement. Earlier in the QuILT, students had already learned that
$\langle \psi_n|\Psi \rangle =\int \psi ^\ast_n (x)\Psi(x)dx$. They had also been asked to differentiate between an energy eigenfunction $\psi_1(x)=\sqrt{\frac{2}{a}}\sin (\frac{\pi x}{a})$ of a 1D infinite square well and a position eigenfunction $ \psi(x)=\delta(x-x_0)$ with eigenvalue $x_0$. In the QuILT, students were explicitly asked to project the ground state of the system $|\psi_1\rangle$ onto the position eigenstate $|x_0\rangle$ with eigenvalue $x_0$ and interpret their result. $|\langle x_0| \psi_1\rangle|^2=|\int\limits^{+\infty}_{-\infty}\delta(x-x_0)\sqrt{2\//a}\sin(\pi x/a)dx|^2=| \sqrt{2\//a}\sin(\pi x_0/a)|^2$ is the probability density for finding the particle at the position $x_0$. Moreover, by the definition of wavefunction, $|\psi_1(x)|^2 =|\langle x| \psi_1\rangle|^2 $ is the probability density for finding the particle at position $x$. The QuILT required students to assimilate the Born interpretation of the probability density for finding the particle with the method of projecting the state vector along a position eigenstate.

After students had learned about the probability density for position measurement using the projection method, the QuILT explicitly brings up a common difficulty they have in differentiating between the probability of obtaining a particular value, the expectation value and similar looking expressions. For example, students are asked to consider the following statement:
\begin{itemize}
    \item If the initial state is $|\Psi\rangle $ for a particle in a 1-D infinite square well, $|\langle \psi_1| H|\Psi\rangle|^2$ is the probability of obtaining energy  when measuring the energy of the particle. Do you agree with this statement? Explain.
\end{itemize}
Students are given hints to consider the dimension (unit) of $\langle \psi_1| H|\Psi\rangle$ . They are also asked to consider the physical meaning of $\langle \Psi| H|\Psi\rangle $ and $\langle \Psi| x|\Psi\rangle$ (in terms of the average of a large number of measurements on identically prepared systems). The warm-up tutorial had already helped students learn that the energy eigenstates $|\psi_n\rangle$ satisfy the TISE $ \hat H |\psi_n\rangle=E_n|\psi_n\rangle$. By decomposing the general state $ | \Psi \rangle $  into a linear superposition of $ |\psi_n \rangle $, students can learn that $\langle \psi_1| H|\Psi\rangle= E_1 \langle \psi_1|\Psi\rangle$ has the dimension of energy. They also contemplate over the fact that the expectation value of the energy in state $|\Psi\rangle$  is the average of a large number of measurements on identically prepared systems, i.e., $\langle \Psi| H|\Psi\rangle = \sum\limits_{n} |A_n|^2 E_n $. In an analogous manner, they interpret the expectation value of position  $\langle \Psi| x|\Psi \rangle $. Explicit comparison of the expectation values with the measurement probabilities is designed to help students distinguish between these related concepts.\\
\\
\textbf{D. Time Development of the System after Measurement}
\\

The second section of the measurement QuILT focuses on the time development of a quantum system after a measurement. After an energy measurement, the system collapses into a stationary state and remains in that state until another measurement is performed. If one were to measure, e.g., the position of the particle, the wavefunction of the system will subsequently evolve in time. In the QuILT, the time-evolution of a quantum system after energy and position measurement were explicitly compared to help students learn about the differences between stationary and non-stationary states.

In the first section of the measurement QuILT, students learn about the possible outcomes of the energy measurement in a 1D infinite square well for three different cases where the states of the system are $|\psi_1\rangle$, $ \frac{1}{\sqrt{2}} (|\psi_1\rangle+|\psi_2\rangle)$ and $\sum\limits_{n} A_n|\psi_n\rangle$ at time $t=0$ when the measurements are performed. At the beginning of the second part of the measurement QuILT, we ask students about the possible values of the energy measurement if we started with the same three initial states but performed the measurement at a time $t>0$. Also, they are explicitly asked to write the states of the system right before the measurement in each case. For example, if the initial state is $|\psi_1\rangle $, the wavefunction at time $t$ would be $|\psi_1\rangle e^{-iE_1t\//\hbar}$ which is still the ground state and the energy measurement will yield the ground state energy $E_1$ with 100\% probability. If the initial state is  $ \frac{1}{\sqrt{2}} (| \psi_1\rangle +\psi_2\rangle)$ the state of the system will evolve into $\frac{1}{\sqrt{2}}(| \psi_1\rangle e^{-iE_1t\//\hbar} +|\psi_2\rangle e^{-iE_2t\//\hbar})$ after a time $t$. Thus, the probability of measuring energy is unchanged (in this case 50\% each for the ground and first excited state energies) even if the system is in a linear superposition of stationary states. Many students correctly predicted that the energy measurement at time $t>0$ would yield the same values $E_1$ and $E_2$ as at time $t=0$ but they incorrectly justified it by saying that the wavefunction after a time $t$ is the same as that at time $t=0$. Students were asked to check their prediction with simulation showing the time evolution of the absolute value of the wavefunction with two energy eigenstate components. After they observed that the shape of the absolute value of the wavefunction changes with time as shown in \textbf{Fig 5}, contrary to their initial prediction, they tried to examine the state of the system at time $t>0$ to resolve the inconsistency between their prediction and their observation.

  \includegraphics[width=1\textwidth]{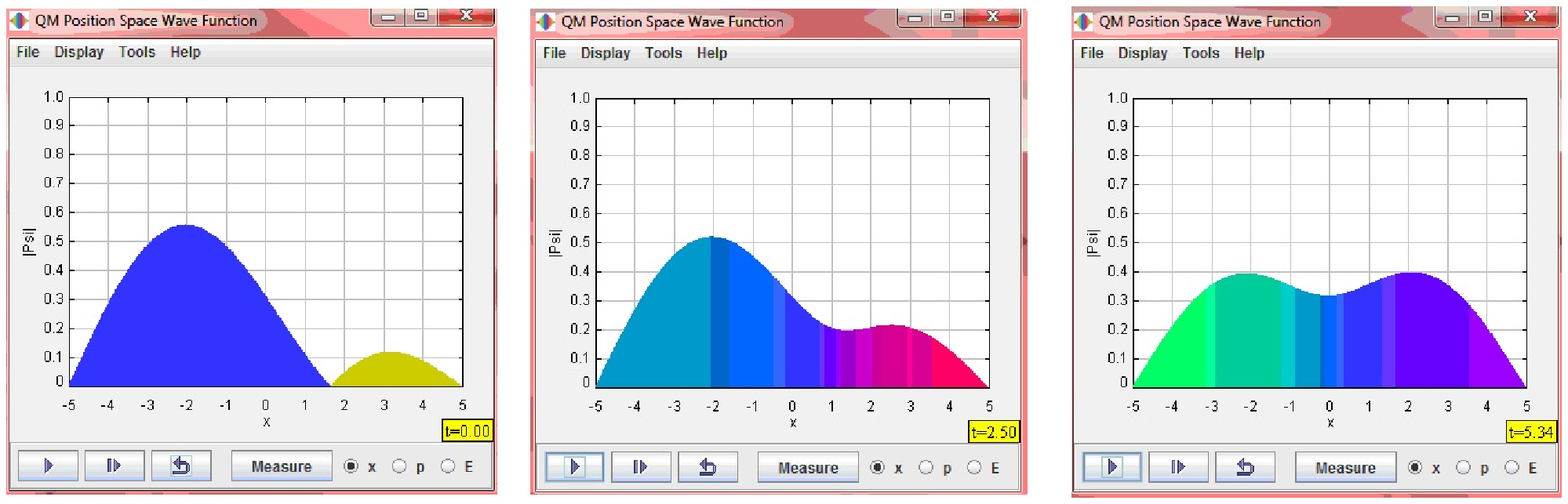}

\textbf{Fig 5}. Time evolution on a superposition state with two energy eigenstate components. (a) is the absolute value of the initial state wavefunction and (b) and (c) are the absolute values of the wavefunction at different times. The phases of the wavefunction are represented by different colors.\\

Students were also asked to repeat the measurement of energy at different time, e.g., t=2 or 3 units after resetting the system to the same initial state after each measurement. They realized that the system only collapsed into $|\psi_1\rangle$ and $ |\psi_2 \rangle $. At this point, the QuILT helped them reason systematically about why the probability of measuring different values of energies does not change with time even though the shape of the wavefunction changes with time for the state $ | \Psi\rangle=\frac{1}{\sqrt{2}}(| \psi_1\rangle e^{-iE_1t\//\hbar} +|\psi_2\rangle e^{-iE_2t\//\hbar})$.

Some students held the misconception that the state of the system after the measurement would eventually go back to the initial state before the measurement. In the QuILT, students observed the time evolution of the wavefunction after the energy measurement and found that the system stays in the stationary state in which it collapsed ($|\psi_1\rangle$ and $|\psi_2 \rangle$) ,as shown in Fig 6(b) and (c) as absolute values in position space, instead of going back to the initial state which is a linear superposition of these states. Students predict and then perform the same sequence of activities with a more general state $|\Psi\rangle=\sum A_n|\psi_n\rangle$ in position space in which more than two coefficients are non-zero. They learn that the wavefunction in this superposition state keeps changing shape with time but the system collapses to one of the energy eigenstates and remains there after the measurement of energy. The QuILT helps the students understand that while the measurement instantaneously collapses the wavefunction, the wavefunction after the measurement evolves in time in a deterministic manner according to the time-dependent Schroedinger equation (TDSE). Moreover, comparison of the time evolution of an energy eigenstate  $|\psi_n\rangle$ (after the measurement) and a general state which is a linear superposition of stationary states (before the measurement) in position space helps build intuition about the meaning of stationary states and non-stationary states.

  \includegraphics[width=1\textwidth]{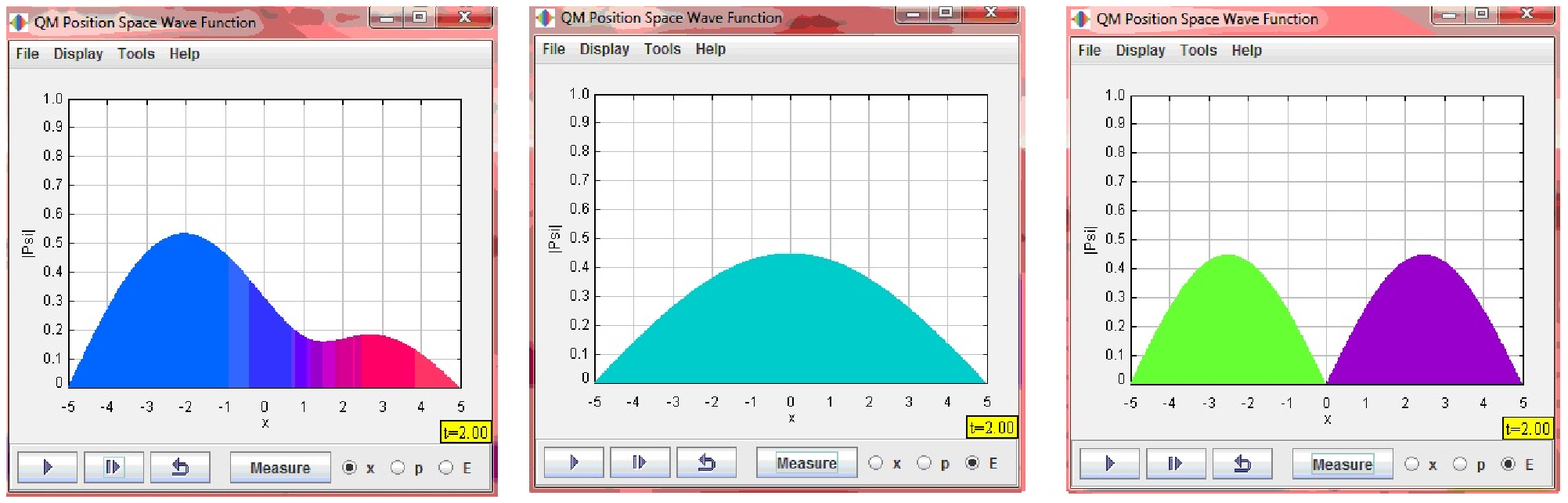}

\textbf{Fig 6}. Energy measurement on a superposition state with two energy eigenstate components after time $t>0$. (a) is the superposition state before the energy measurement. (b) and (c) are the two possible outcomes of the quantum measurement. \\
Many students held the misconception that, after the position measurement, the position eigenstate does not change with time and the system is stuck in a position eigenstate. In the QuILT, students are asked to use the simulation after their initial prediction of what should happen when they perform a position measurement starting from a general state. In an ideal measurement, at the instant the position is measured, the wavefunction of the system will collapse to a delta function $\delta(x-x_0)$ as shown in \textbf{Fig 7(a)}. The position eigenfunction can be written as a linear superposition of the energy eigenfunctions, i.e., $\Psi(x,t=0 )=\delta(x-x_0)=\sum\limits_n A_n \psi_n(x) $. Different energy eigenstates will have their own time-dependent phase factors and the wavefunction $\Psi(x,t)$ would not be a delta function $\delta(x-x_0)$ except at some special times. Fig 7(b) and (c) show selected snapshots of the time evolution of the absolute value of the position eigenfunction.

  \includegraphics[width=1\textwidth]{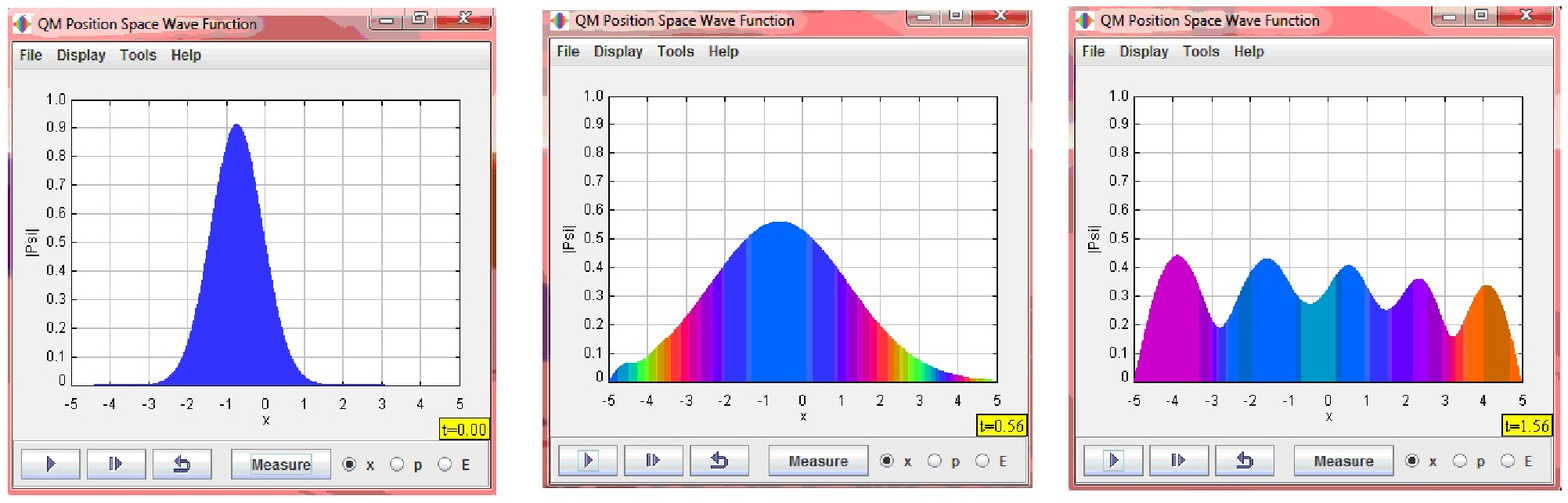}

\textbf{Fig 7}. Time evolution of the position eigenfunction. (a) is the position eigenfunction right after a position measurement. (b) and (c) are the wavefunction of the system at later times after the position measurement.\\

Besides the pictorial representation in the simulations, the QuILT helps students learn to interpret the time evolution of wavefunction via the TDSE and discern the central role of the Hamiltonian of the system in the evolution. The following is an example of a question that students are asked:
\begin{itemize}
    \item Given the wavefunction at time $t=0$, why is it useful to write the state of a quantum system as a superposition of energy eigenstates to find the wavefunction after time $t$?
\end{itemize}
Students must realize that the Hamiltonian governs the time evolution of the system according to the TDSE so the eigenstates of the Hamiltonian are special for issues related to the time evolution of the wavefunction. Help is provided at the end of the QuILT if students are struggling with these issues.

Though students learn formally that the position eigenstate is not a stationary state, some of them still held the misconception that the position eigenstate after a position measurement would finally return to the initial state, e.g., $(|\psi_1\rangle +| \psi_2\rangle)\//\sqrt{2}$. The simulation is helpful in confronting this mistaken belief. The students observe that the delta function does not remain a delta function as shown in \textbf{Fig 7} (although there is revival of the delta function periodically for a 1D infinite square well). They perform a systematic analysis of the time-dependence of wavefunction starting with a delta function to convince themselves that the state will never go back to the state right before the measurement, i.e., $(|\psi_1 \rangle +| \psi_2\rangle)\//\sqrt{2}$. Since the delta function  $\delta(x-x_0)$ contains non-zero coefficients $A_n$ for higher energy eigenstate wavefunction $\psi_n(x) \ \ \  (n>2) $, the probability of measuring these higher energies $|A_ne^{-iE_nt\//\hbar}| ^2$ would never be zero. Therefore, the system cannot return to the state $(|\psi_1\rangle +| \psi_2\rangle)\//\sqrt{2}$ after the position measurement, no matter how long the wait.

It is important that students learn whether the probability of obtaining different energies or positions change with time. For a general wavefunction $\Psi(x,t)=\sum\limits_n A_n \psi_n e^{-iE_nt\//\hbar} $ at time $t$, the probability of obtaining $E_n$ in an energy measurement is a constant $|A_n e^{-iE_nt\//\hbar}|^2=|A_n|^2$ independent of time. However, when position is measured, the probability of finding the particle at $x=x_0$ is $|\Psi(x,t)|^2=|\sum\limits_n A_n \psi_n(x) e^{-iE_nt\//\hbar}|^2 $ which depends on time. This non-trivial time-dependence of the probability of position measurement can be observed in the simulation since the shape of wavefunction keeps changing with time. The QuILT helps students learn to distinguish between the time-dependence of the probability of measuring energy and position through a concrete example of the wavefunction $\frac{1}{ \sqrt{2}}[\psi_1(x)e^{-iE_1t\//\hbar}+\psi_2(x)e^{-iE_2t\//\hbar}]$.\\
\\
\textbf{E. Consecutive Measurements  }
\\

After students learned about how to analyze the time evolution of the wavefunction according to the TDSE after the measurement of a physical observable, the related concepts were reinforced by asking them questions about consecutive measurements. For example, students were asked about the possible outcomes of an energy measurement after a position measurement in the state
$\frac{1}{\sqrt{2}}(|\psi_1\rangle +| \psi_2\rangle)$. Some students incorrectly claimed that one can only obtain energies $E_1$ or $E_2$. However, since the position measurement will collapse the system to a position eigenstate which is a superposition of the energy eigenstates $|\psi_n\rangle$ (including those corresponding to very high energies), the energy measurement that follows the position measurement could yield a very high value $E_n$. After the prediction, students use the simulation in position space to check their prediction and find that the wavefunction could collapse to an energy eigenstate $\psi_n$ with $n\geq3$  as shown in \textbf{Fig 8}. Students are also asked to calculate the probability for measuring different energy values.

  \includegraphics[width=1\textwidth]{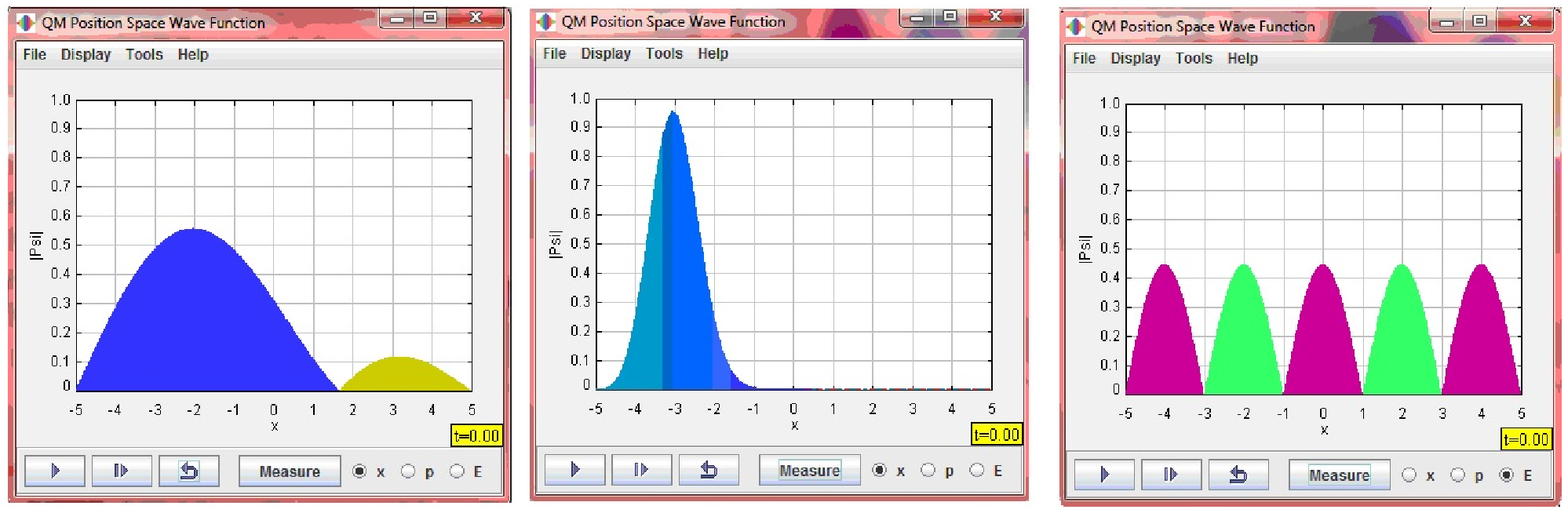}

\textbf{Fig 8}. Energy measurement after a position measurement of the initial state with only two energy eigenstate components n=1 and n=2 as shown in (a). Following the position measurement in (b), the energy measurement yields the energy eigenvalue with n=5 as shown in (c).\\

  \includegraphics[width=1\textwidth]{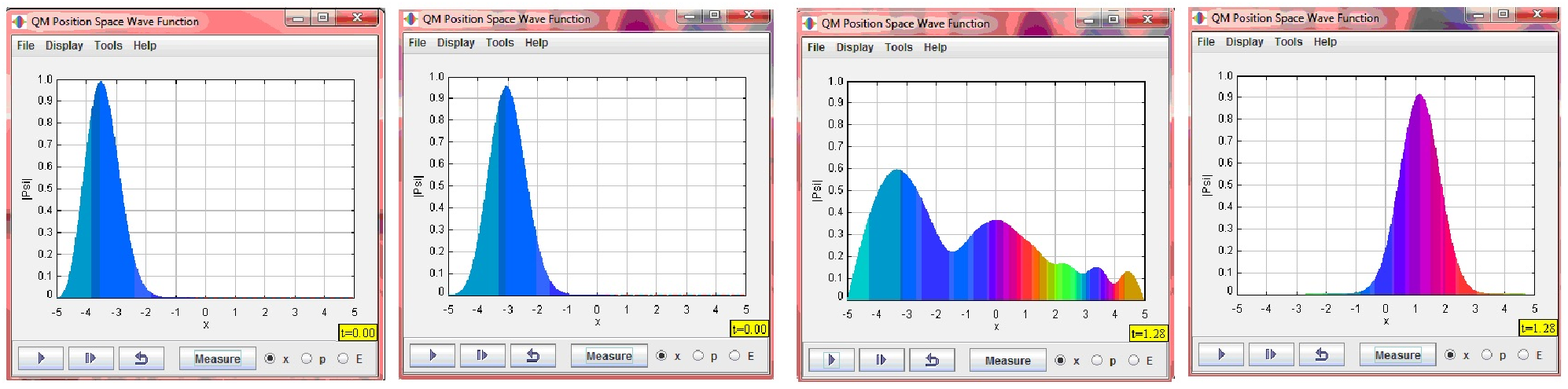}

\textbf{Fig 9}. Consecutive position measurement in quick succession (a and b) /after waiting for some time (c and d)\\

Students are also asked to predict what would happen if they made two consecutive position measurements quickly so that the wavefunction does not have the time to evolve. In the simulation, they find that when the second measurement was made immediately after the first measurement, the particle is found approximately at the same position since the wavefunction does not have the time to evolve. On the other hand, the wavefunction would not be a peaked delta function if we waited for some time before performing the second measurement and we can find the particle at a different position. The simulation provides the flexibility of stopping or starting the time evolution at any point (or even stepping through time-evolution slowly) so that students can note the differences between the consecutive position measurements performed in quick succession as shown in \textbf{Fig 9(a)} and \textbf{(b)} vs. slowly as shown in \textbf{Fig 9(c)} and \textbf{(d)}.

\section{PRELIMINARY EVALUATION}

\ \ \ \ We designed a pre-test and a post-test to assess some issues related to measurement after the traditional instruction, after concept tests related to measurement (pre-test) and after working on the measurement QuILT (post-test). To eliminate any possible differences in the difficulty levels of the pre-test and the post-test, we divided the tests into two versions, i.e., Test A and Test B. Test A and Test B both had two multiple choice questions (Q1 and Q2) and four open-ended questions (Q3-Q6). We randomly assigned Tests A and B when we distributed the pre-test and post-test to students in both the comparison group and the experimental group. In the experimental group, students who obtained Test A in the pre-test were given Test B in the post-test and vice versa.

The comparison group students only had traditional lectures in class and regular homework problems from the textbook. Students in the comparison group took the test at the end of their first semester quantum mechanics course when all the topics about quantum measurement had been taught. The class average (25 students) was 26\% including both Test A and Test B. The experimental group students had been using the concept tests as a peer instruction tool in class since the first day of the semester. The pre-test was given to the students after the lecture and the average score for 31 students was 76\%. The experimental group students were given the QuILT as homework after being administered the pre-test in class. When they turned in the QuILT as homework, they were administered the post-test. Their post-test average score was 91\% for 29 students (2 students absent in the post-test).

To analyze students' understanding of different concepts and principles in quantum measurement, we calculated the percentage of correct responses for each question in Test A and Test B as listed in Table 1 and Table 2. The numbers in the brackets represent the number of students who answered that question. The concepts involved in each question are also shown in the Tables.\\
\\
\\
\\
\\

\begin{table}
  \centering
  \caption{ The pre-test and post-test scores on each question of Test A. The concepts involved in each question are shown in italic.}
  \begin{tabular}{|c|c|c|c|}
  \hline
 Test A  & Comparison & Experimental Group &Experimental Group  \\
         & Group (15) & Pre-test (15)      &Post-test (15)    \\   \hline
     & Traditional  & Lecture \& Concept  & Lecture \& Concept  \\
     & Lecture Only &  Test             & Test \& QuILT        \\     \hline
   Q1&\multicolumn{3}{|c|}{\emph{whether a wavefunction is an energy eigenstate}} \\ \cline{2-4}
     &13\%  &67\% & 87\% \\  \cline{2-4}\hline
   Q2&\multicolumn{3}{|c|}{\emph{energy measurement outcomes of a superposition state}}  \\   \cline{2-4}
     &40\%  &93\% & 100\% \\  \cline{2-4}\hline
   Q3&\multicolumn{3}{|c|}{\emph{sketch the shape of a position eigenstate and find the probability}} \\  \cline{2-4}
     &37\%  &77\% & 87\% \\  \cline{2-4}\hline
   Q4&\multicolumn{3}{|c|}{\emph{probability of energy measurement}}  \\     \cline{2-4}
     &3\%   &62\% & 77\% \\  \cline{2-4}\hline
   Q5&\multicolumn{3}{|c|}{\emph{consecutive position measurement after position measurement}} \\    \cline{2-4}
     &37\%  &78\% & 100\% \\  \cline{2-4}\hline
   Q6&\multicolumn{3}{|c|}{\emph{consecutive energy measurement after energy measurement}} \\        \cline{2-4}
     &53\%  &70\% & 93\% \\  \cline{2-4}
  \hline
\end{tabular}
\end{table}

From Q1 in Test A and Q2 in Test B, we observe that the concept tests in class resolved many of the students' difficulties related to the difference between the stationary states and the eigenstates of the operators corresponding to any physical observables. Q4 in Test A and Q3 in Test B suggest that after the research-based learning tools, students can better apply the projection method to calculate the probabilities of measuring a physical observable. When answering Q3 in Test A and Q4 in Test B after using the concept tests and the QuILT, students also showed improved interpretation of the shapes of the eigenfunctions for different operators corresponding to different physical observables. For the questions related to the time development of the wavefunction after a measurement, e.g., Q6 in both Test A and B, the QuILT led students to a better performance compared to when they used the concept tests only. Also, the QuILT helped the students have an improved understanding of the measurement outcome and the properties of the corresponding eigenstate. After learning the QuILT, more students could correctly answer the questions related to consecutive measurements such as Q5 in both Test A and B. Due to the limitation of sample size, individual performance might affect the average score on each question. However, the effectiveness of the concept tests and QuILTs in improving students' performance is reflected by the difference in the overall performance of the experimental group and the comparison group.

\begin{center}
\begin{table}[h]
  \centering
  \caption{The pre-test and post-test scores on each question of Test B. The concepts involved in each question are shown in italic.}
\begin{tabular}{|c|c|c|c|}
 \hline
 Test B  & Comparison & Experimental Group &Experimental Group  \\
         & Group (10) & Pre-test (16)      &Post-test (14)    \\   \hline
     & Traditional  & Lecture \& Concept  & Lecture \& Concept  \\
     & Lecture Only &  Test             & Test \& QuILT        \\     \hline
   Q1&\multicolumn{3}{|c|}{\emph{what state will the system be in after a quantum measurement}} \\ \cline{2-4}
     &50\%  &69\% & 86\% \\  \cline{2-4} \hline
   Q2&\multicolumn{3}{|c|}{\emph{what is a stationary state}}  \\ \cline{2-4}
     &0\%   &75\% &  79\% \\  \cline{2-4}\hline
   Q3&\multicolumn{3}{|c|}{\emph{energy measurement outcomes and probabilities}} \\  \cline{2-4}
     &15\%  &97\% & 100\% \\  \cline{2-4}\hline
   Q4&\multicolumn{3}{|c|}{\emph{sketch the shape of an energy eigenstate}}  \\   \cline{2-4}
     &35\%  &88\% & 96\% \\  \cline{2-4}\hline
   Q5&\multicolumn{3}{|c|}{\emph{consecutive position measurement after energy measurement}} \\    \cline{2-4}
     &10\%  &66\% & 89\% \\  \cline{2-4}\hline
   Q6&\multicolumn{3}{|c|}{\emph{consecutive energy measurement after position measurement}} \\     \cline{2-4}
     &0\%   &66\% & 93\% \\  \cline{2-4}
  \hline
\end{tabular}
\end{table}
\end{center}

\section{SUMMARY AND CONCLUSION}

\ \ \ \ \ Students have common difficulties in learning the issues related to quantum measurement. We have developed the research-based learning tools such as the QuILT and concept tests to improve students' understanding of quantum measurement concepts. Both these learning tools keep students actively engaged in the learning process. They provide a guided approach to bridge the gap between the quantitative and conceptual issues related to quantum measurement, help students connect different concepts and build a knowledge structure. Our preliminary results show that these learning tools are effective in improving students' understanding about quantum measurement.

\begin{acknowledgments}
This material is based upon work supported by the National Science
Foundation.
\end{acknowledgments}


\begin{thebibliography}{99}
\addcontentsline{toc}{section}{References}
\bibitem{add} G. Zhu and C. Singh, preceding article, Improving students?
understanding of quantum measurement. II.
Development of research-based learning tools, Phys.
Rev. ST Phys. Educ. Res. \textbf{8}, 010117 (2012).
\bibitem{1} E. Mazur, \emph{Peer Instruction: A User's Manual}. (Prentice Hall, Upper Saddle River, NJ, 1997).
\bibitem{2}C. Crouch and E. Mazur, Peer Instruction: Ten years of experience and results, Am. J. Phys. \textbf{69}, 970 (2001)
\bibitem{3}McDermott and the Physics Education Group at the University of Washington. \emph{Physics by Inquiry, Vols. I and II}. (John Wiley \& Sons Inc., New York, NY, 1996)
\bibitem{4}C. Singh, Interactive learning tutorials on quantum mechanics, Am. J. Phys. \textbf{76}, 400 (2008).
\bibitem{5}C. Singh, Impact of peer interaction on conceptual test performance, Am. J. Phys. \textbf{73}, 446 (2005).
\bibitem{6}D. P. Maloney, T. L. O'Kuma, C. J. Hieggelke, and A. V. Heuvelen, Surveying students' conceptual knowledge of electricity and magnetism, Am. J. Phys. \textbf{69}, S12 (2001).
\bibitem{7}G. Zhu and C. Singh, Improving students¡¯ understanding of quantum mechanics via the Stern-Gerlach experiment, Am. J. Phys. \textbf{79}, 499 (2011)
\bibitem{8}N. W. Reay, P. Li, and L. Bao, Testing a new voting machine question methodology, Am. J. Phys. \textbf{76}, 171 (2008).
\bibitem{9}M. Chi, Thinking Aloud, in \emph{The Think Aloud Method: A Practical Guide to Modeling Cognitive Processes}, edited by M. W. Van Someren, Y. F. Barnard, and J. A. C. Sandberg (Academic, London, 1994).
\bibitem{10}J. Hiller, I. Johnston, D. Styer, \emph{Quantum Mechanics Simulations, Consortium for Undergraduate Physics Software}, (John Wiley and Sons, New York, 1995); D. Schroeder and T. Moore, A computer-simulated Stern-Gerlach laboratory, Am. J. Phys. \textbf{61}, 798 (2003).
\bibitem{11}D. Zollman, S. Rebello, and K. Hogg, Quantum physics for everyone: Hands-on activities integrated with technology, Am. J. Phys. \textbf{70}, 252 (2002); P. Jolly, D. Zollman, S. Rebello, and A. Dimitrova, Visualizing potential energy diagrams, Am. J. Phys. \textbf{66}, 57 (1998).
\bibitem{12}C. E. Wieman, K. K. Perkins, and W. K. Adams, Oersted medal lecture: Interactive simulations for teaching physics: What works, what doesn't, and why, Am. J. Phys. \textbf{76} (4\&5), 393 (2008).
\bibitem{13}S. B. McKagan, K. Perkins and C. Wieman, Reforming large lecture modern physics course for engineering majors using a PER-based design, in \emph{2006 Physics Education Research Conference Proceedings}, edited by L. McCullough, P. Heron, and L. Hsu (AIP Press, Melville, NY, 2007), p. 34.
\bibitem{14}R. N. Steinberg, G. E. Oberem, and L. C. McDermott, Development of a computer-based tutorial on the photoelectric effect, Am. J. Phys. \textbf{64}, 1370 (1996).
\bibitem{15}For example, see http://www.opensourcephsyics.org,
\bibitem{16}M. Belloni, W. Christian and A. Cox, \emph{Physlet Quantum Physics}, Pearson Prentice Hall, Upper Saddle River, NJ, (2006). M. Belloni and W. Christian, Physlet for quantum mechanics, Comput. Sci. Eng. \textbf{5}, 90 (2003).
\bibitem{17} Leslie Smith, Making Education Sense of Piaget's Psychology, Oxford Review of Education \textbf{11}, 181-191 (1985).
\end{thebibliography}
\end{document}